\documentclass[aps,showpacs,floats,superscriptaddress]{revtex4-1}

\usepackage[dvips]{graphicx}
\usepackage{amssymb}
\usepackage{amsmath,bm}
\usepackage{psfrag}
\usepackage{epsfig}
\usepackage{float}

\begin{document}
\title{Analysis of a combined influence of substrate wetting and surface electromigration on a thin film stability and dynamical morphologies}

\author{Mikhail Khenner}
\affiliation{Department of Mathematics, Western Kentucky University, Bowling Green, KY 42101}

\begin{abstract}
A PDE-based model combining surface electromigration and wetting is developed for the analysis of morphological stability of ultrathin solid films.
Adatom mobility is assumed anisotropic, and two directions of the electric field (parallel and perpendicular to the surface) are discussed and contrasted. 
Linear stability analyses of small-slope evolution equations are performed, followed by computations of fully nonlinear parametric evolution equations
that permit surface overhangs. The results reveal parameter domains of instability for wetting and non-wetting films and variable electric field strength, 
nonlinear steady-state solutions in certain cases, and interesting coarsening behavior for strongly wetting films.

\vskip 0.5\baselineskip

\noindent
{\it Submitted to the special issue "Nanoscale wetting of solids on solids" of the journal Comptes Rendus Physique (Olivier Pierre-Louis, Univ. Lyon, Editor)}

\vskip 0.5\baselineskip
  \vskip 0.5\baselineskip

\end{abstract}
\date{\today}
\maketitle

\section{Introduction}
\label{Intro}

Since the early 1990's there has been an interest in understanding and accessing the effects of electromigration on kinetic instabilities 
of crystal steps \cite{St}-\cite{UCS} and epitaxial islands \cite{KD}-\cite{KKHV}, and on morphological stability of epitaxial films 
\cite{BMOPL}-\cite{TGM1}.
Among other applications (which were developed primarily at the microscale), electromigration has been also used for fabrication of nanometer-sized 
gaps in metallic films. Such gaps are suitable for testing of the conductive properties of single molecules and control of their functionalities \cite{VFDMSKBM}-\cite{GSPBT}. 
For instance, Ref. \cite{GSPBT} describes fabrication of nanoscale contacts by using electromigration to thin down and finally break the epitaxially grown ultrathin
(10 ML) Ag films wetting the Si(001) substrate. The gap between contacts can be cyclically opened and closed by applying electromigration current at 80 K to open the gap, and enabling surface diffusion  by annealing to the room temperature, to close it.

Thus for this and other emerging applications at the nanoscale \cite{WH,SV} it seems important to understand and characterize the effects of substate 
wetting and electromigration that are simulaneously active in the physical system. This paper combines these effects in a model that is based on 
an evolution equation(s) for the continuous profile of the film surface. The focus is on wetting films with isotropic surface energy, but with anisotropic adatom mobility \cite{KD,SK}, although
the model allows any combination of wetting properties and anisotropies. We factor in and discuss the effects on film stability and morphological evolution of
the electric field that is either parallel, or perpendicular to the initial planar surface of the film, and do not limit considerations to small deviations from planarity, i.e.
the arbitrary surface slopes and even surface overhangs are permitted by the model. Models of wetting appropriate for continuum-level modeling of the 
surface diffusion-based dynamics of solid films
have been developed and discussed extensively primarily in the context of thin film heteroepitaxy \cite{CG}-\cite{K2}. Our analysis is based on one such model, called the 
two-layer exponential model for the surface energy \cite{ORS,GolovinPRB2004,LevinePRB2007}, \cite{GillWang}-\cite{BWA} (which is particularly useful when the surface energy is anisotropic), 
but other models of wetting discussed in Refs. \cite{CG}-\cite{K2} can be used instead, and the results are expected to be qualitatively similar.
The goal of modeling in this paper is not to match the theoretical results to the experiment \cite{GSPBT} and thus help in understanding the experiment, but rather to provide the broad analysis of the interplay of two effects (wetting and electromigration) that, at least to our knowledge, has not been addressed in prior publications.

\section{Problem Statement}
\label{ProbStat}

We consider a 2D single-crystal film of unperturbed height $H_0$ with the 1D parametric surface\\ $\Gamma(x(u,t),z(u,t))$, where
$x$ and $z$ are the Cartesian coordinates of a point on a surface, $t$ is the time and $u$ is the parameter along the surface.
The origin of the Cartesian reference frame is on the substrate, and along the substrate (the $x$-direction) the film is
assumed infinite. The $z$-axis is in the direction normal
to the substrate and to the initial planar film surface.
Surface marker particles will be used for computations of surface dynamics \cite{Trygg}. Thus $x$ and $z$ ($z>0$)
in fact represent the coordinates of
a marker particle, which are governed by the two coupled parabolic PDEs \cite{Sethian1,BKL}:
%
\begin{eqnarray}
x_t &=& V z_s = V \frac{1}{g}z_u, \label{base_eq1}\\
z_t &=& - V x_s = - V \frac{1}{g}x_u \label{base_eq2}.
\end{eqnarray}
Here (and below) the subscripts $t,s,u,z$ and $x$ denote partial differentiation with respect to these variables,
$V$ is the normal velocity of the surface which incorporates the physics of the problem,
and $g(u,t)=s_u=z_u/\cos{\theta}=\sqrt{x_u^2+z_u^2}$. Here $s$ is the surface arclength and $\theta$ the surface orientation angle, i.e. the one that the unit surface normal makes with the reference crystalline direction (chosen along the $z$-axis).
If the surface slopes are bounded at all times (surface does not overhang), then it is more convenient to describe surface dynamics by a
single evolution PDE for the height function $h(x,t)$ of the film. 
Eqs. (\ref{base_eq1}), (\ref{base_eq2}) can be easily reduced to such ``h-equation", which we will use for analysis; however, Eqs. (\ref{base_eq1}), (\ref{base_eq2}) will be used for most computations
in this paper. Similar parametric approach was used recently in Ref. \cite{OL} for the computation of a hill-and-valley structure coarsening in the presence of material deposition (growth)
and strongly anisotropic surface energy. 

Assuming that temperature is sufficiently high and surface diffusion is operative, the normal velocity is given by
\begin{equation}
V= \frac{D\Omega \nu}{kT}\left[\left\{M(\theta)\mu_s\right\}_s + \alpha E_0q\left\{M(\theta)f(\theta)\right\}_s\right],
\label{1.4c}
\end{equation}
where $D$ is the adatoms diffusivity, $\Omega$ the atomic volume, $\nu$ the adatoms surface density, $kT$ the Boltzmann's factor,
$\mu$ the surface chemical potential, $M(\theta)$ the anisotropic adatom mobility, $E_0$ the applied electric field, $q>0$ the effective charge of adatoms,
$f(\theta)=\sin{\theta}$, if the electric field is vertical, or $f(\theta)=\cos{\theta}$, if the electric field is horizontal, and 
$\alpha=\pm 1$ is used to select either stabilizing, or destabilizing action of the field for the chosen combination of the vertical or horizontal orientation of the field and the mobility $M(\theta)$.
The two values of $\alpha$ correspond to two possible field orientations once either the horizontal, or the vertical field direction has been chosen (that is, field directed up-down, or left-right).
The first term in $V$ describes the high-T surface diffusion, the second term describes surface diffusion enabled by electromigration \cite{M}.

The surface chemical potential $\mu$ is assumed to have the contributions from the surface energy and the surface wetting interaction with the substrate/film interface:
\begin{equation}
\mu = \Omega \left[\gamma\kappa+\gamma_z\cos{\theta}\right],
\label{1.4e}
\end{equation}
where $\kappa=\theta_s=g^{-3}\left(z_{uu}x_u-x_{uu}z_u\right)$ is the surface mean curvature and, in the general case, $\gamma(z,\theta)$ is the height- and orientation-dependent, i.e. anisotropic, surface energy. (Note again that $z$ here stands for the shortest distance between the 
substrate and a chosen point $(x,z)$ on a film surface; this distance is the height $h(x,t)$ of the surface if there is no overhangs.) In this paper we focus on the effects due to anisotropic adatom mobility, thus we will use the simpler \textit{isotropic}  model for the surface energy 
\cite{CG}-\cite{K2}:
\begin{equation}
\gamma=\gamma(z)  = \gamma_f + \left(\gamma_S-\gamma_f\right)\exp{\left(-z/\ell\right)},
\label{1.4fp}
\end{equation}
where $\gamma_S$ is the (constant) energy of a substrate/gas (or vacuum) interface,  $\ell$ the characteristic wetting length, and $\gamma_f$ the constant energy of a crystal/gas interface (that is, of the film surface).
Eq. (\ref{1.4fp}) is the interpolation between the two energies. In the limit of a thick film, $z\rightarrow \infty$,
only the latter energy is retained in this expression (because the inter-molecular forces between the substrate and the surface molecules are relatively short-ranged), and in the limit of a film of zero thickness only the former energy is retained. Despite that Eq. (\ref{1.4fp}) is phenomenological, it matches surprisingly well the experiments and the ab-initio calculations (at least for lattice-mismatched systems) \cite{GillWang,WKSM,BWA}.

\begin{figure}[h]
\centering
\includegraphics[width=2.0in]{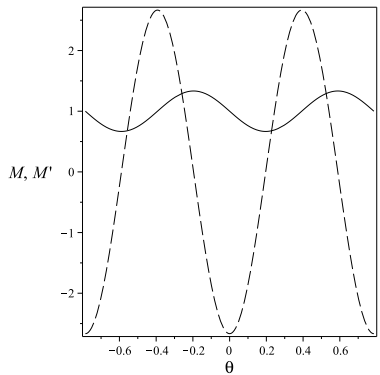}
\caption{Anisotropic mobility $M(\theta)$ (solid line) and its derivative (dashed line) for $\beta=1,\ N=4$ and $\phi=\pi/16$.}
\label{Fig_Mobility}
\end{figure}

Finally, we assume the (dimensionless) anisotropic adatom mobility in the form \cite{SK}
\begin{equation}
M(\theta)= \frac{1+\beta\cos^2{[N(\theta+\phi)]}}{1+\beta\cos^2{[N\phi]}}, \label{Mobility}
\end{equation}
where $N$ is the number of symmetry axes and $\phi$ is the angle between a symmetry direction and the average surface orientation. $\beta$ is a parameter determining the strength of the anisotropy.
Throughout the paper we present results either for $\beta=0$ (isotropic case), or for $\beta=1$ and $N=4,\; \phi=\pi/16$. For the latter set of parameters values the graphs of the functions 
$M(\theta)$ and $M'(\theta)$ are shown in Fig. \ref{Fig_Mobility}.

\textit{Remark 1.} In the limit $z\rightarrow \infty$ (where wetting effect is not operative), the present problem is most closely related to the one analyzed 
by Schimschak and Krug in Ref. \cite{SK}. The essential difference is that these authors determine the electric field from the solution of the potential equation with the appropriate boundary
conditions on the material boundaries, including the moving surface \cite{KAIN,B}. Thus their solution for the electric field is nonlocal, unlike the local approximation used in this paper. 
The local nature of the electric field explains why we did not detect traveling wave solutions, which are the hallmark of Refs. \cite{SK,B}. 

Next, we choose $\ell$ as the length scale and $\ell^2/D$ as the time scale and write the dimensionless counterparts of Eqs. (\ref{base_eq1}), (\ref{base_eq2}), where we use same notations for dimensionless variables: 
\begin{eqnarray}
x_t &=& \left[B\left\{M(\theta)\mu_s\right\}_s + A\left\{M(\theta)f(\theta)\right\}_s\right] z_s \equiv V z_s, \label{base_eq1_1}\\
z_t &=& - V x_s \label{base_eq2_1},\\
\mu &=& \left[1+(G-1)\exp{(-z)}\right]\kappa+(1-G)\exp{(-z)}\cos{\theta},\; \mbox{where}\; \cos{\theta} = z_s. \label{base_eq3_1}
\end{eqnarray}
Notice that the dimensionless forms of $f(\theta)$ and the parametric expression for $\kappa$ coincide with their dimensional forms, and that Eq. (\ref{1.4fp}) has been substituted in Eq. (\ref{1.4e}). For conciseness, we keep differentiation with respect to the arclength, $s$, in the dimensionless equations, but their most transparent forms for the computations result when the 
differentiation with respect to $s$ is replaced with the differentiation with respect to the parameter $u$, using $\partial/\partial s = (1/g)\partial/\partial u$. In Eqs. (\ref{base_eq1_1}) - (\ref{base_eq3_1}) $B=\Omega^2\nu \gamma_f/\left(kT\ell^2\right)$ is the surface diffusion parameter, $A=\alpha \nu \Omega E_0 q/(kT)$ is the strength of the electric field, and $G=\gamma_S/\gamma_f$ is the ratio
of substrate to film surface energies. For wetting films $G>1$, for non-wetting films $0<G<1$. Notice that $A$ may take on positive or negative values through the parameter $\alpha$.

As was mentioned above, if there is no overhangs then Eqs. (\ref{base_eq1_1})-(\ref{base_eq3_1}) can be reduced to the single dimensionless evolution equation for the surface height $h$:
\begin{equation}
h_t = B\left[M\left(h_x\right)\left(1+h_x^2\right)^{-1/2}\mu_x\right]_x + A\left[M\left(h_x\right)\left(1+h_x^2\right)^{-1/2}f\left(h_x\right) \right]_x, \label{h-eq}
\end{equation}
where 
\begin{equation}
\mu = \left[1+(G-1)\exp{(-h)}\right]\kappa+(1-G)\exp{(-h)}\left(1+h_x^2\right)^{-1/2},\quad \kappa = -h_{xx}\left(1+h_x^2\right)^{-3/2},
\end{equation}
$f\left(h_x\right) = h_x$, if the electric field is vertical, or $f\left(h_x\right) = 1$, if the electric field is horizontal. Also $M\left(h_x\right)$ is given by Eq. (\ref{Mobility}),
where $\theta$ is replaced by $arctan\left(h_x\right)$.  

Using values: $D=1.5\times 10^{-6}$ cm$^2$/s, $\Omega = 2\times10^{-23}$ cm$^3$, $\gamma_f = 2\times10^3$ erg/cm$^2$,
$\nu = 10^{15}$ cm$^{-2}$, $kT = 1.12\times10^{-13}$ erg, $\ell= 3\times 10^{-8}$ cm (0.3 nm, or 1ML), gives $B=8$. Using $q=5e$ (where $e=5\times 10^{-10}$ statcoulombs is the absolute value of the electron charge)
and the minimum applied voltage difference $\Delta V = 5\times 10^{-3}$ Volts acting across the typical distance of the order of the film height, $\Delta L = 10$ nm,  gives $|qE_0| = 4\times 10^{-4}$ erg/cm, which translates to $|A| = 71$. Applied voltage in surface electromigration experiments at the nanoscale can be as high as 1 V \cite{GSPBT}, thus in this study we explored the range of field strengths $71\le |A|\le 7100$.

In the following sections we analyze several representative situations.

\section{Vertical electric field}
\label{VertField}

\subsection{Linear stability analysis}
\label{VertField_Linear}

The small slope approximation $\left(|h_x|\ll 1\right)$ of Eq. (\ref{h-eq}) reads:
\[
h_t = BM(0)\left[\left\{(1-G)\exp(-h) - 1\right\}h_{xxx}+h_x (G-1)\exp(-h)\left(h_{xx}+1\right)\right]_x +  
\]
\begin{equation}
\hspace{0.7cm} BM'(0)(G-1)\left[\exp(-h)h_x^2\right]_x + AM(0)h_{xx} - \frac{3}{2}AM(0)h_x^2h_{xx}+ 2AM'(0)h_x h_{xx}, \label{h-eq-ssa-vert}
\end{equation}
where the mobility has been linearized about the flat surface $h_x=0$, that is, $M\left(h_x\right) = M(0)+M'(0)h_x,\; M(0)>0$. The last two terms are the
simplest nonlinearities from the expansion of the electromigration flux that involve $M(0)$ and $M'(0)$. Without loss of generality
we will take $M(0)=1$ here and elsewhere in this paper. (See Fig. \ref{Fig_Mobility}. When mobility is isotropic, i.e. $\beta=0$, then $M=M(0)=1$, as is seen from Eq. (\ref{Mobility}).)
Notice that the anisotropy of the mobility, $M'(0)$, does not have an effect on linear stability, as it enters in 
the coefficients of the nonlinear terms. 
Also note that the last three terms can be written in a conservative form similar to the terms in the first line of the equation and to the first term in the second line
(and this is how they are implemented in the code).

Introducing small perturbation $\xi(x,t)$ in Eq. (\ref{h-eq-ssa-vert}) by replacing $h$ with $h_0+\xi(x,t)$ (where $h_0 = H_0/\ell$ is the dimensionless unperturbed film height), linearizing
in $\xi$ and assuming normal modes for $\xi$, gives the perturbation growth rate $\omega(k)$, where $k$ is the perturbation wavenumber:
\begin{equation}
\omega(k) = -B\left[1+(G-1)\exp(-h_0)\right]k^4 - \left[B(G-1)\exp(-h_0)+A \right]k^2 \label{grrate-vert}.
\end{equation}

\textit{Remark 2.} When wetting interaction is absent (thick film: $h_0\rightarrow \infty$), Eq. (\ref{grrate-vert}) reduces to the standard one, $\omega(k) = -Bk^4 - Ak^2$, which reflects the stabilizing action of the surface diffusion
and either stabilizing ($A>0$, electric field is in the positive $z$ direction), or destabilizing ($A<0$, electric field is in the negative $z$ direction) action of the electric field. 
Such film is absolutely stable when $A>0$, but when $A<0$ it is long wave-unstable.  

\subsubsection{Analysis of Eq. (\ref{grrate-vert})}
\label{VertField_Linear_analysis}
\medskip
\begin{itemize}

\item \textit{Wetting films} $(G>1)$. From Eq. (\ref{grrate-vert}) one notices that wetting films are absolutely linearly stable when $A>0$, but they are
long wave-unstable when $A<-B(G-1)\exp(-h_0)<0$ (see Fig. \ref{longwave_inst}). The short-wavelength cut-off wavenumber, the maximum growth rate, and the wavenumber at which the latter
occurs are: 
\begin{equation}
k_c = \sqrt{\frac{-A-B(G-1)\exp(-h_0)}{B[1+(G-1)\exp(-h_0)]}},\quad \omega_{max} = \frac{1}{4}\frac{\left[A+B(G-1)\exp(-h_0)\right]^2}{B[1+(G-1)\exp(-h_0)]},\quad k_{max} = \frac{k_c}{\sqrt{2}}.
\label{kc_etc}
\end{equation}
$k_c$ and $\omega_{max}$ are plotted in Figs. \ref{Fig_kc_vs_h0_A_G} and  \ref{Fig_omegamax_vs_h0_A_G}. The film stability \textit{decreases} with increasing $h_0$, and this trend saturates around $h_0=10$ (3 nm). That is, for the stated field strength $A=-71$ films of thickness $h_0 > 10$ do not "feel" the stabilizing presence of the substrate, and they are as stable as the films that do not interact with the substrate at all.
Of course, increasing field strength $|A|$ makes the film less stable, but increasing $G$ makes it more stable, since the substrate energy provides stabilizing effect. 
In Figs. \ref{Fig_kc_vs_h0_A_G}(c) and \ref{Fig_omegamax_vs_h0_A_G}(c) the rate of decrease of $k_c$ and $\omega_{max}$  with $G$ (i.e., the rate of stabilization) increases fast with decreasing $h_0$.
For $A=-71$ at $h_0 \sim 3.5$ and $G \sim 300$ the entire dispersion curve $\omega(k)$ is below the $k$-axis, see the dashed curves in Figs. \ref{Fig_kc_vs_h0_A_G}(c) and \ref{Fig_omegamax_vs_h0_A_G}(c),
and thus the film is absolutely linearly stable for $h_0 <\sim 3.5$ and $G >\sim 300$. By analysing the condition for longwave instability, $A<-B(G-1)\exp(-h_0)<0$, one can get an understanding of why this happens.
\begin{figure}[h]
\vspace{-2.5cm}
\centering
\includegraphics[width=4.0in]{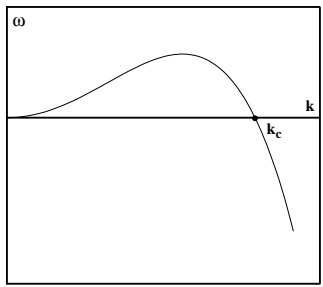}
\vspace{-6.5cm}
\caption{Sketch of perturbation's linear growth rate $\omega(k)$ corresponding to longwave instability of the film surface.}
\label{longwave_inst}
\end{figure}
\begin{figure}[h]
\centering
\includegraphics[width=2.0in]{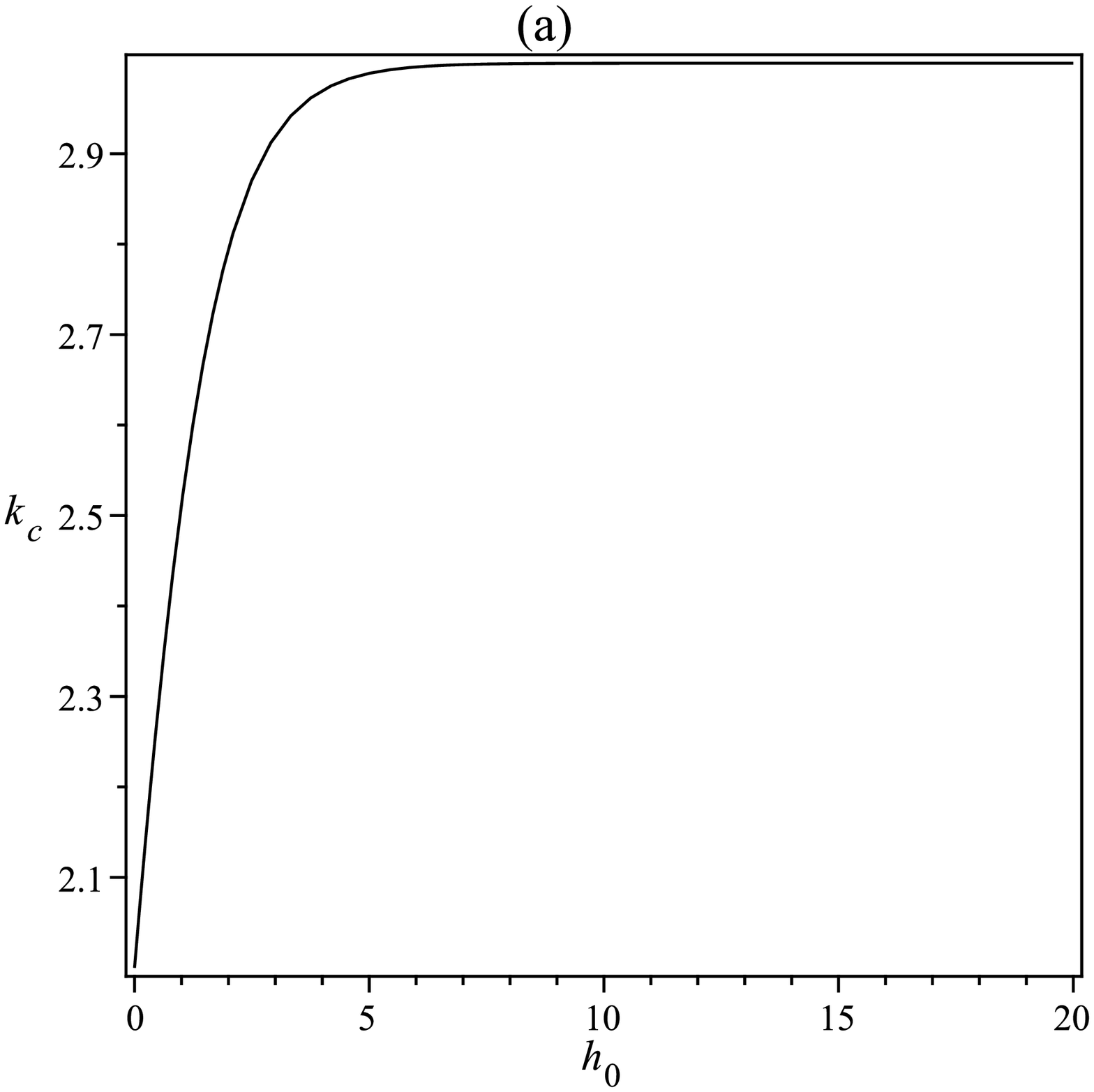}\includegraphics[width=2.0in]{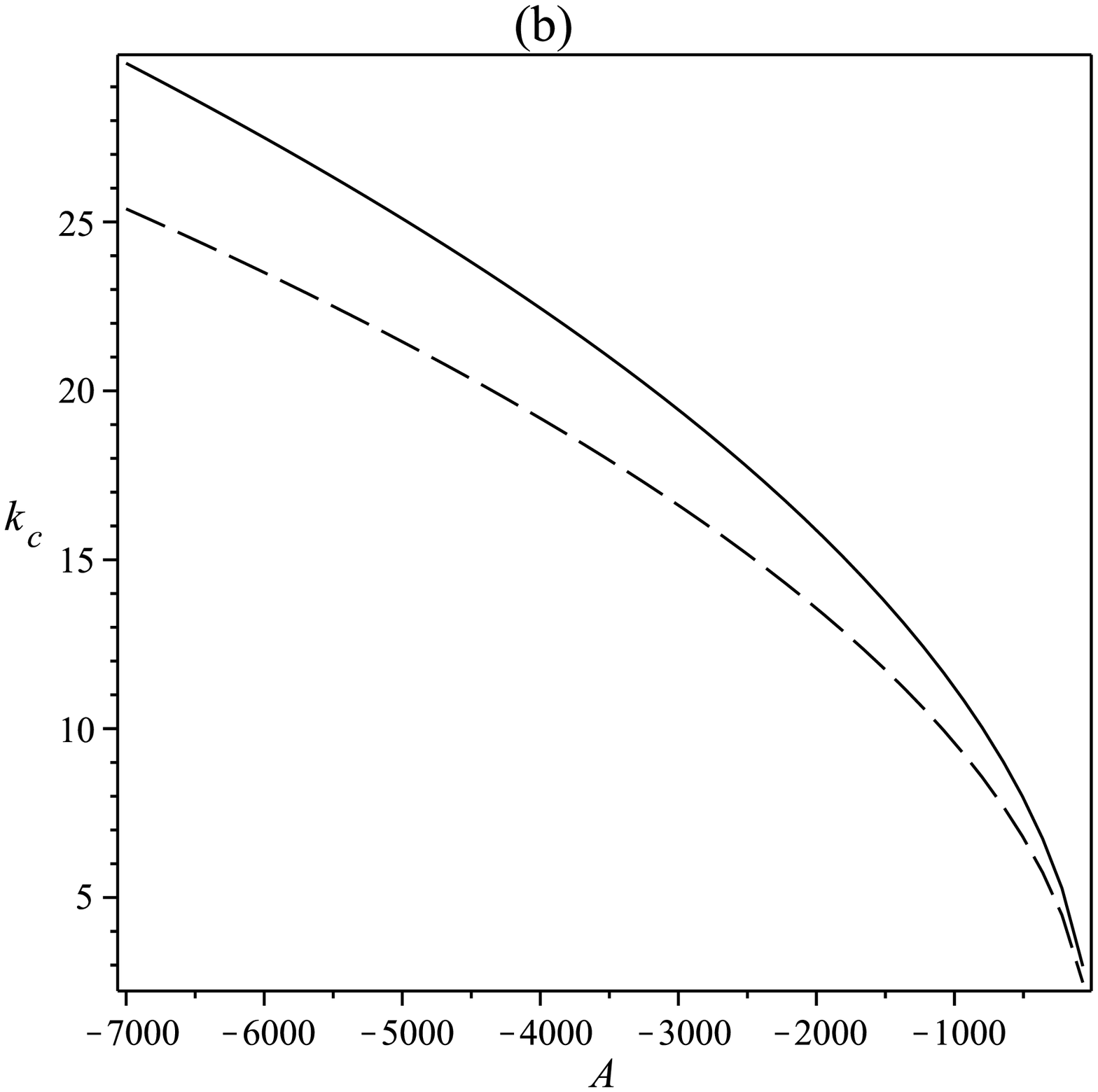}\includegraphics[width=2.0in]{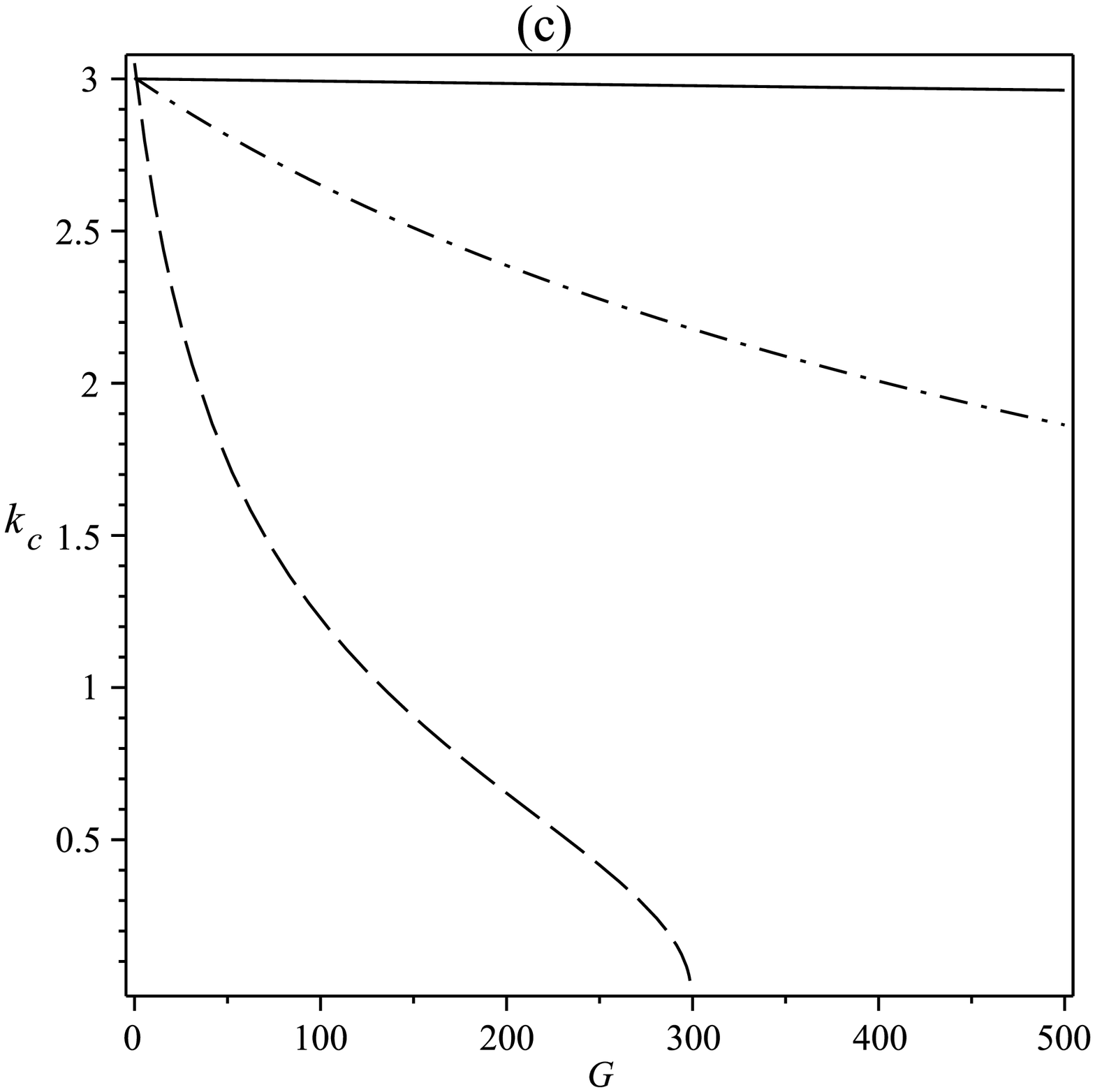}
\caption{Wetting films and vertical electric field: Instability cut-off wavenumber $k_c$. (a) vs. $h_0$, $A=-71,\ G=2$; (b) vs. $A$, $h_0=10$ (solid line), $h_0=1$ (dashed line). $G=2$; (c) vs. $G$, 
$h_0=10$ (solid line), $h_0=6$ (dash-dotted line), $h_0=3.5$ (dashed line). $A=-71$.}
\label{Fig_kc_vs_h0_A_G}
\end{figure}
\begin{figure}[h]
\centering
\includegraphics[width=2.0in]{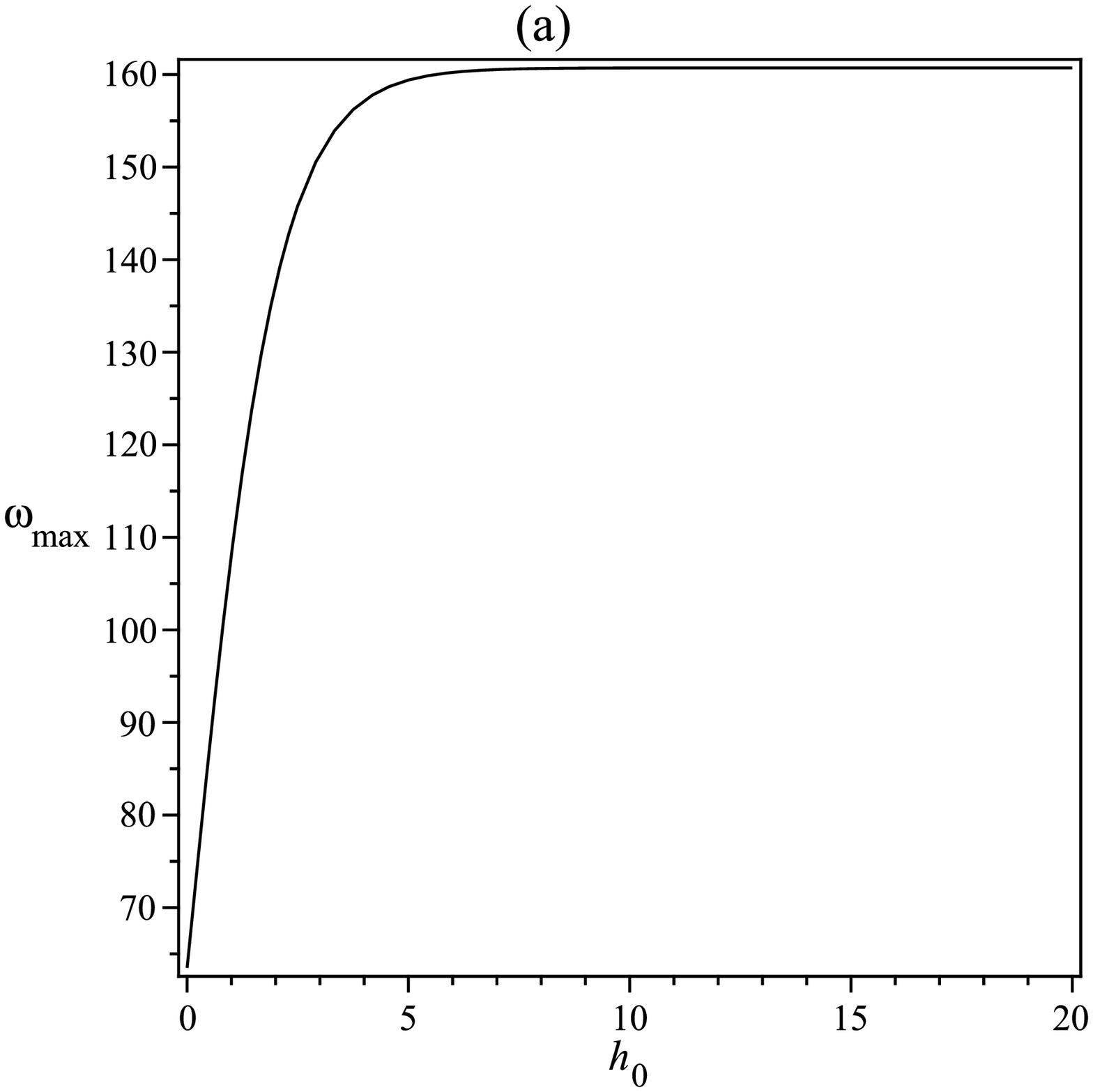}\includegraphics[width=2.0in]{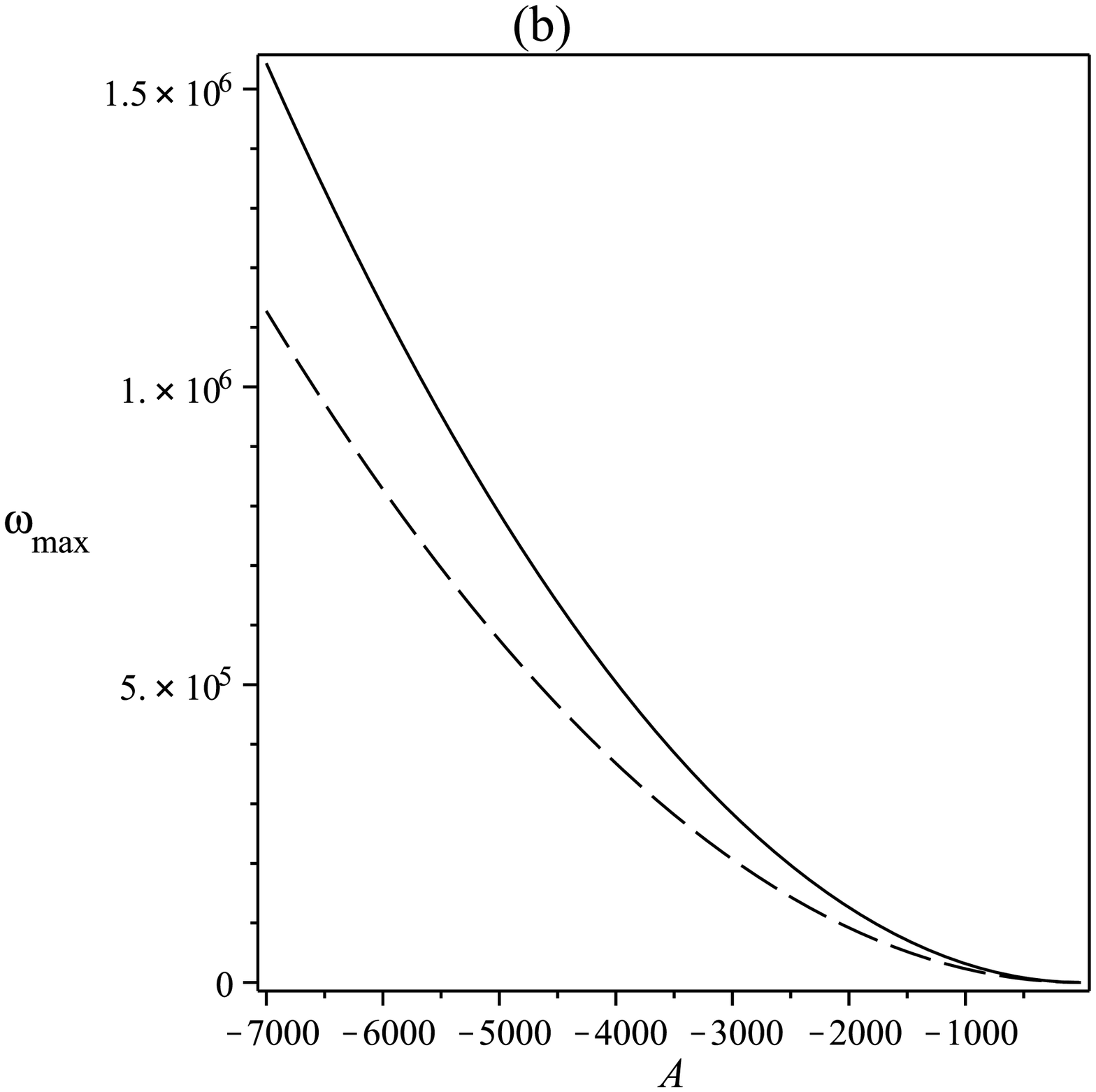}\includegraphics[width=2.0in]{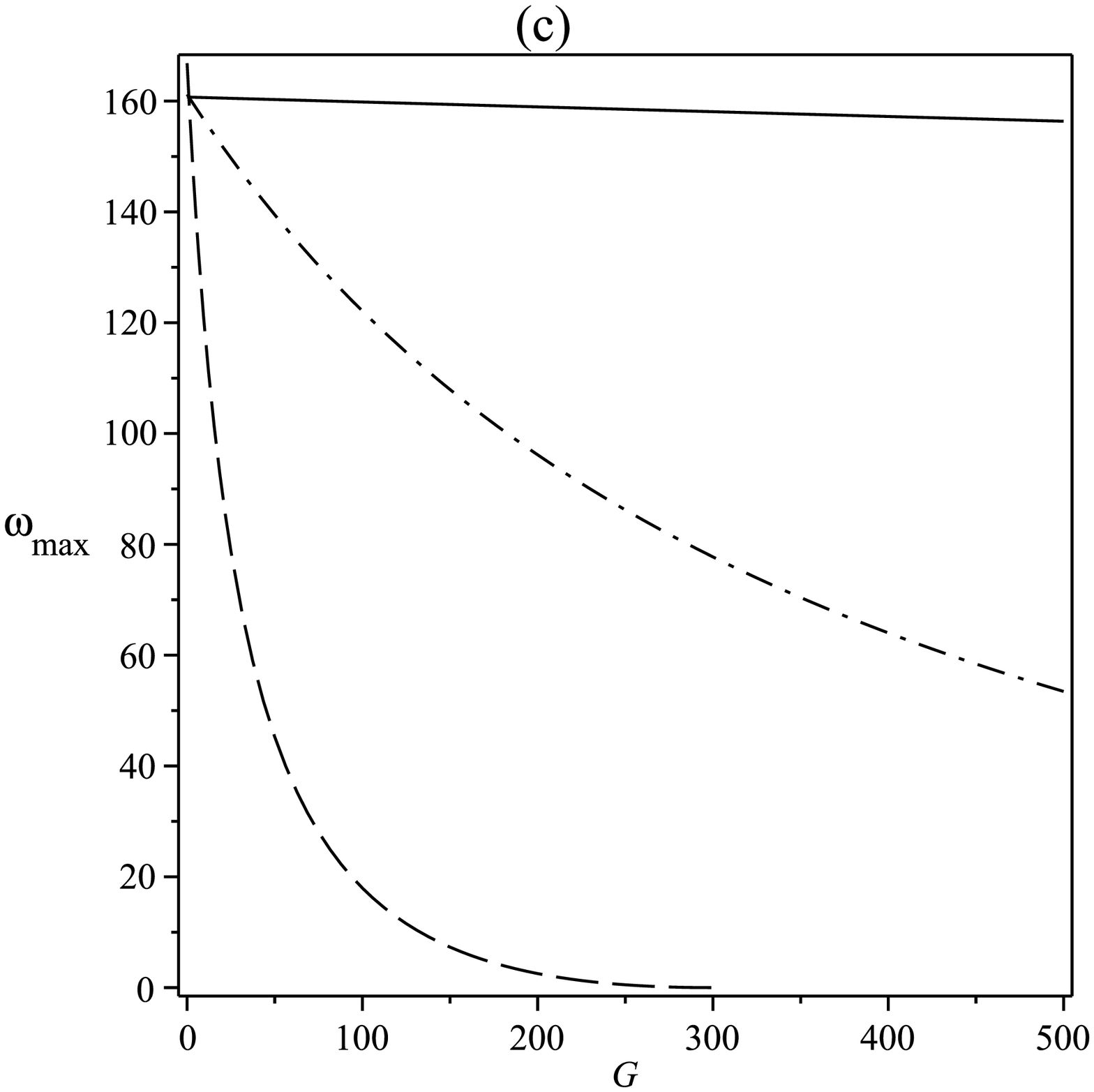}
\caption{Wetting films and vertical electric field: Maximum growth rate $\omega$. Cases (a)-(c) are as in Fig. \ref{Fig_kc_vs_h0_A_G}.}
\label{Fig_omegamax_vs_h0_A_G}
\end{figure}
Indeed, this condition is equivalent to $h_0 > \ln{\left(BA^{-1}(1-G)\right)}$, and for the typical values of $A$ and $B$ stated above and for moderately large $G$ the right-hand side of the latter inequality is negative, thus
the longwave instability occurs for any film height $h_0$. This is similar to the situation when wetting effects are absent, see \textit{Remark 2} (but of course the spectrum of unstable wavenumbers and the magnitude of the maximum growth rate are different). There is a critical thickness $h_{0c}$ below which the film is absolutely linearly stable only if $G-1 \sim |A|$, i.e. when $G$ is of the order
of at least one hundred. 

\medskip
\item \textit{Non-wetting films} $(0<G<1)$. 
In this case, for $A<0$ the long-wave instability occurs for any film thickness and any electric field strength - which is again similar to the situation when wetting effects are absent, see \textit{Remark 2}. When $A>0$ the long-wave instability (with $k_c,\; \omega_{max}$ and $k_{max}$ numerically similar to those shown in Eq. (\ref{kc_etc})) 
occurs when the film height is less than critical, 
$h_0 < h_{0c} = \ln{\left(AB^{-1}(1-G)^{-1}\right)}$. When $h_0 > h_{0c}$ the film is absolutely linearly stable. $h_{0c}$ is plotted in Fig. \ref{Fig_h0crit_vs_A_G}. It is clear that
$h_{0c}$ increases with increasing either $A$, or $G$, or both parameters.

\begin{figure}[h]
\vspace{-5.3cm}
\centering
\includegraphics[width=5.0in]{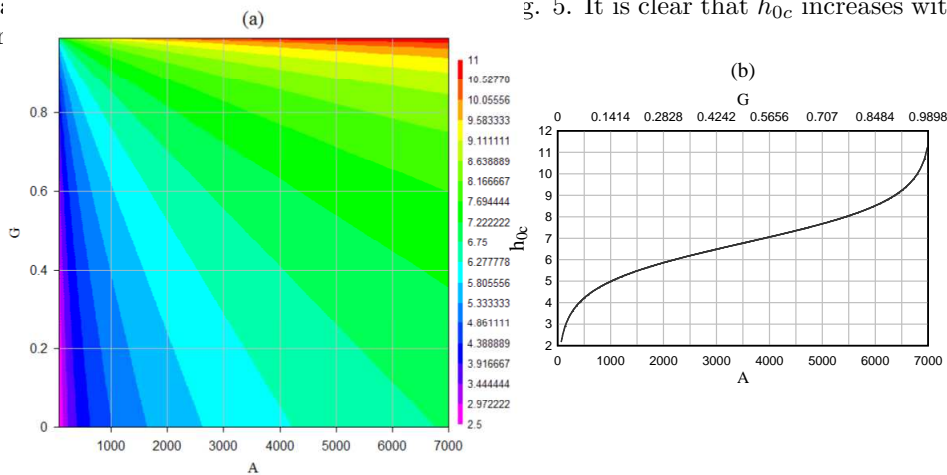}
\vspace{-6.0cm}
\caption{(Color online). Non-wetting films and vertical electric field: (a) Contour plot of the critical thickness $h_{0c}$. (b) Diagonal cross-section (from the lower left corner to the upper right corner)  
of the contour plot in (a).}
\label{Fig_h0crit_vs_A_G}
\end{figure}

\end{itemize}

\subsection{Nonlinear surface dynamics of wetting films}

\begin{figure}[h]
\vspace{-1.5cm}
\centering
\includegraphics[width=2.5in]{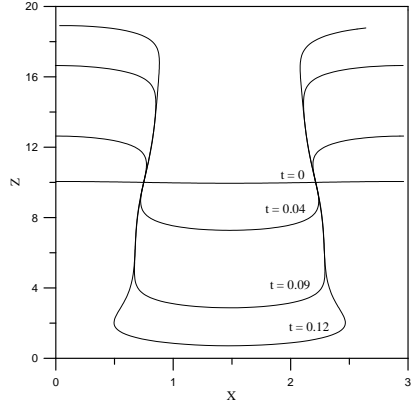}
\vspace{-1.2cm}
\caption{Wetting films, vertical electric field, isotropic mobility: One wavelength of the surface morphology on the periodic domain $0\le x\le \lambda_{max}$, computed using Eq. (\ref{h-eq-ssa-vert}) 
($M'(0)=0$) starting from the small-amplitude cosine-shaped initial condition. $A=-71,\ h_0=10,\ G=2$.}
\label{Fig_VertField_ConstMob_ic=1}
\end{figure}

Computations with the small-slope Eq. (\ref{h-eq-ssa-vert}) (where $M(0)=1,\; M'(0)=0$) were performed for 
$G=2$ and varying field strengths $A$ that satisfy the condition for long-wave instability, $A<-B(G-1)\exp(-h_0)<0$. Computational domain was $0\le x\le \lambda_{max},\;  \lambda_{max} = 2\pi/k_{max}$ with periodic boundary conditions, and the initial condition was 
the small-amplitude cosine-shaped perturbation of the flat surface $h_0=const.$ 
All computations produced steady-state solutions that have the shape of a vertically stretched cosine curve with a fairly large amplitude. However, neither of these steady-states were confirmed when instead fully nonlinear parametric equations  (\ref{base_eq1_1}) - (\ref{base_eq3_1}) were computed.\footnote{Of course, we first carefully checked that values of $k_c,\ k_{max}$
and $\omega_{max}$ from the linear stability analyses are reproduced in the computations of the parametric equations. This is one of the methods that we used for checking that the parametric code is errors free.} 
Fig. \ref{Fig_VertField_ConstMob_ic=1} shows the computed morphology. Overhangs are clearly visible, and the bottom of the "pit" flattens out as it approaches the substrate due to increase of the repulsive 
force, promoting overhangs in the vicinity. 

\begin{figure}[h]
\vspace{-4.5cm}
\centering
\includegraphics[width=3.5in]{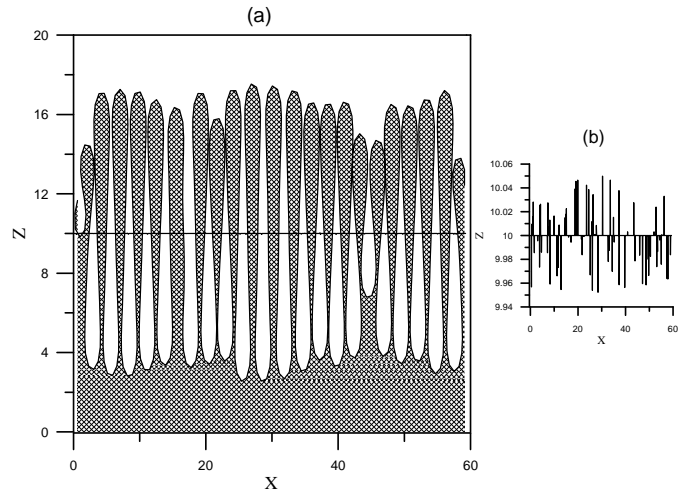}
\vspace{-0.5cm}
\caption{Wetting films, vertical electric field, isotropic mobility: 
(a) Surface morphology computed using Eqs. (\ref{base_eq1_1}) - (\ref{base_eq3_1}) ($\beta=0$ in Eq. (\ref{Mobility})) on the domain $0\le x\le 20\lambda_{max}$, starting from the
small-amplitude random initial condition shown in (b). $A=-71,\ h_0=10,\ G=2$.}
\label{Fig_VertField_ConstMob_ic=3}
\end{figure}

Similarly, when Eq. (\ref{h-eq-ssa-vert}) was computed with the random small-amplitude initial condition on the domain $0\le x\le 20\lambda_{max}$ (the mobility was again isotropic), the result was a perpetually coarsening
hill-and-valley structure. This was again not confirmed in the computations of parametric equations. The late time morphology computed using the parametric equations is shown in Fig. \ref{Fig_VertField_ConstMob_ic=3}(a). The surface
develops deep ``pockets" whose walls overhang and eventually merge, resulting in tear drop-shaped voids trapped in the solid; this can be seen, for instance, on the interval $40<x<50$. 
Fig. \ref{Fig_VertField_ConstMob_ic=3}(b) shows the typical random initial condition.

Anisotropic mobility resulted in morphologies that are similar to ones shown in Figs. \ref{Fig_VertField_ConstMob_ic=1} and \ref{Fig_VertField_ConstMob_ic=3}(a), but skewed left or right, 
depending on the sign of $\phi$. As we pointed in Sec. \ref{VertField_Linear}, anisotropy matters in the nonlinear stage of the dynamics.

\section{Horizontal electric field}
\label{HorField}

\subsection{Linear stability analysis}
\label{Hor_Linear}

In the case of the horizontal electric field the small slope approximation of Eq. (\ref{h-eq}) reads:
\[
h_t = B\left[\left\{(1-G)\exp(-h) - 1\right\}h_{xxx}+h_x (G-1)\exp(-h)\left(h_{xx}+1\right)\right]_x +  
\]
\begin{equation}
\hspace{0.7cm} BM'(0)(G-1)\left[\exp(-h)h_x^2\right]_x - Ah_xh_{xx} + AM'(0)h_{xx} - \frac{3}{2}AM'(0)h_x^2h_{xx}. \label{h-eq-ssa-hor}
\end{equation}
Again here we retained two simplest nonlinearities from the expansion of the electromigration flux term. The coefficient $M(0)$ in the third-to-last term has been set equal to one. This
nonlinear term and the last (nonlinear) term in Eq. (\ref{h-eq-ssa-hor}) do not have an effect on linear stability.
Unless mobility is isotropic, the second-to-last term of Eq. (\ref{h-eq-ssa-hor}) has an effect on linear stability.
Assuming anisotropy, the growth rate is
\begin{equation}
\omega(k) = -B\left[1+(G-1)\exp(-h_0)\right]k^4 - \left[B(G-1)\exp(-h_0)+A M'(0)\right]k^2 \label{grrate-hor}.
\end{equation}
Comparing Eqs. (\ref{grrate-vert}) and (\ref{grrate-hor}), it is clear that one can obtain stability properties in the horizontal field case by replacing
$A$ in Eq. (\ref{grrate-vert}) with $AM'(0)$.
Since $M'(0)\approx -2.7$ (see Fig. \ref{Fig_Mobility}) for the chosen parameter values in Eq. (\ref{Mobility}), then for instability one must have 
$A>0$ (the electric field is in the positive $x$ direction). Thus, for instance, the stability analysis of wetting films in the paragraph 1. of Sec. \ref{VertField_Linear_analysis} translates directly into the case of the horizontal field simply by replacing $A$ 
in the formulas by $-2.7A$, where $A>0$.  Figs. \ref{Fig_kc_vs_h0_A_G}(b) and \ref{Fig_omegamax_vs_h0_A_G}(b) are also valid in the case of the horizontal field if the
$A$ values along the horizontal axes are understood as values of the product $-2.7A$ (thus the absolute characteristic $A$ values are 
roughly three times smaller than in the case of the vertical field).

\subsection{Nonlinear surface dynamics of wetting films}

\subsubsection{Periodic steady-states from sinusoidal perturbations}
\medskip

\begin{figure}[h]
\vspace{-2.0cm}
\centering
\includegraphics[width=3.5in]{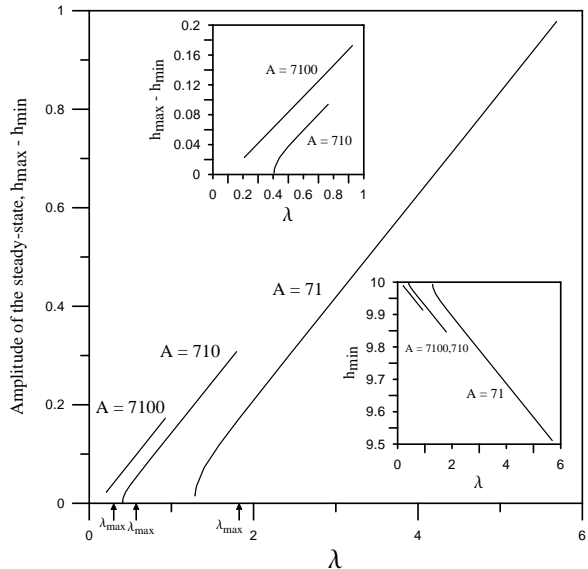}
\vspace{-1.5cm}
\caption{Wetting films, horizontal electric field, anisotropic mobility: Amplitude of the steady state vs. the perturbation wavelength $\lambda$ for different strengths of the electric field. 
Upper inset - zoom into the area of the graph around $\lambda=0.5$.
Lower inset - minimum height of the steady state profile vs. $\lambda$, also for different strengths of the electric field. Arrows labeled ``$\lambda_{max}$" point out 
the $\lambda_{max}$ value for each of the three electric field strengths. Parameters values are $h_0=10,\ G=2$.}
\label{Fig_horizontal_ic=1}
\end{figure}

Computations with either the small-slope Eq. (\ref{h-eq-ssa-hor}), or the 
parametric equations (\ref{base_eq1_1})-(\ref{base_eq3_1}), of the evolution of a one wavelength ($\lambda$)\footnote{In this computation $\lambda$ is not necessarily equal to $\lambda_{max}$.}, small-amplitude cosine curve-shaped perturbation on a periodic domain resulted in steady-state profiles which have a shape of a vertically stretched cosine curve
with a fairly large amplitude. The steady-state profiles often displayed a more sharply peaked bottom and less curved walls than the cosine curve, and the amplitude is significantly smaller in the (fully nonlinear) parametric case. In fact, it can be noticed from Fig. \ref{Fig_horizontal_ic=1} that the film is nowhere close to dewetting the substrate for all tested field strengths, notwithstanding that the stabilizing influence of the substrate is minimal for the chosen parameters values (see the discussion in Sections \ref{VertField_Linear_analysis}, paragraph 1, and \ref{Hor_Linear}).

\textit{Remark 3.}  When Eq. (\ref{h-eq-ssa-hor}) is used in a computation, the last term there is at large responsible for the saturation of the surface slope and the existence of the steady state. When this nonlinear term is omitted from the equation, the slope increases until the computation breaks down.

In the computations of the parametric equations, unless $\lambda$ is larger than approximately $4\lambda_{max}$, the surface evolves towards the steady-state mostly by vertical stretching of the initial 
shape. Perturbations of larger wavelength develop a large-amplitude, hill-and-valley type distortions which slowly coarsen into a steady-state, cosine curve-type shape. 
Fig. \ref{Fig_horizontal_ic=1} shows the amplitude of the steady-state profile, $h_{max}-h_{min}$, and the height of the profile at it's lowest point, $h_{min}$, vs. $\lambda$.
It can be seen that, at least for moderate strengths of the electric field, growth of the amplitude is logarithmic in the vicinity of $\lambda_c$, and for $\lambda\gg \lambda_c$ growth is linear. 
The numerically determined slope of the linear section of $A=71$ curve is 0.208, the one of the $A=710$ curve is 0.217, and the one of the $A=7100$ curve is 0.209, which suggests that 
the slope is insensitive (or very weakly sensitive) to the field strength.
The decay of $h_{min}$ with increasing wavelength mirrors the growth of the amplitude, thus the steady-state surface shape is symmetrically vertically stretched about the equilibrium
surface position $h_0=10$. All computed steady-states are stable with respect to imposition of random small- and large-amplitude point perturbations, which we confirmed by 
computing the dynamics of such perturbed shapes.

\subsubsection{Coarsening of random initial roughness}
\medskip

\begin{figure}[h]
\vspace{-1.2cm}
\centering
\includegraphics[width=3.0in]{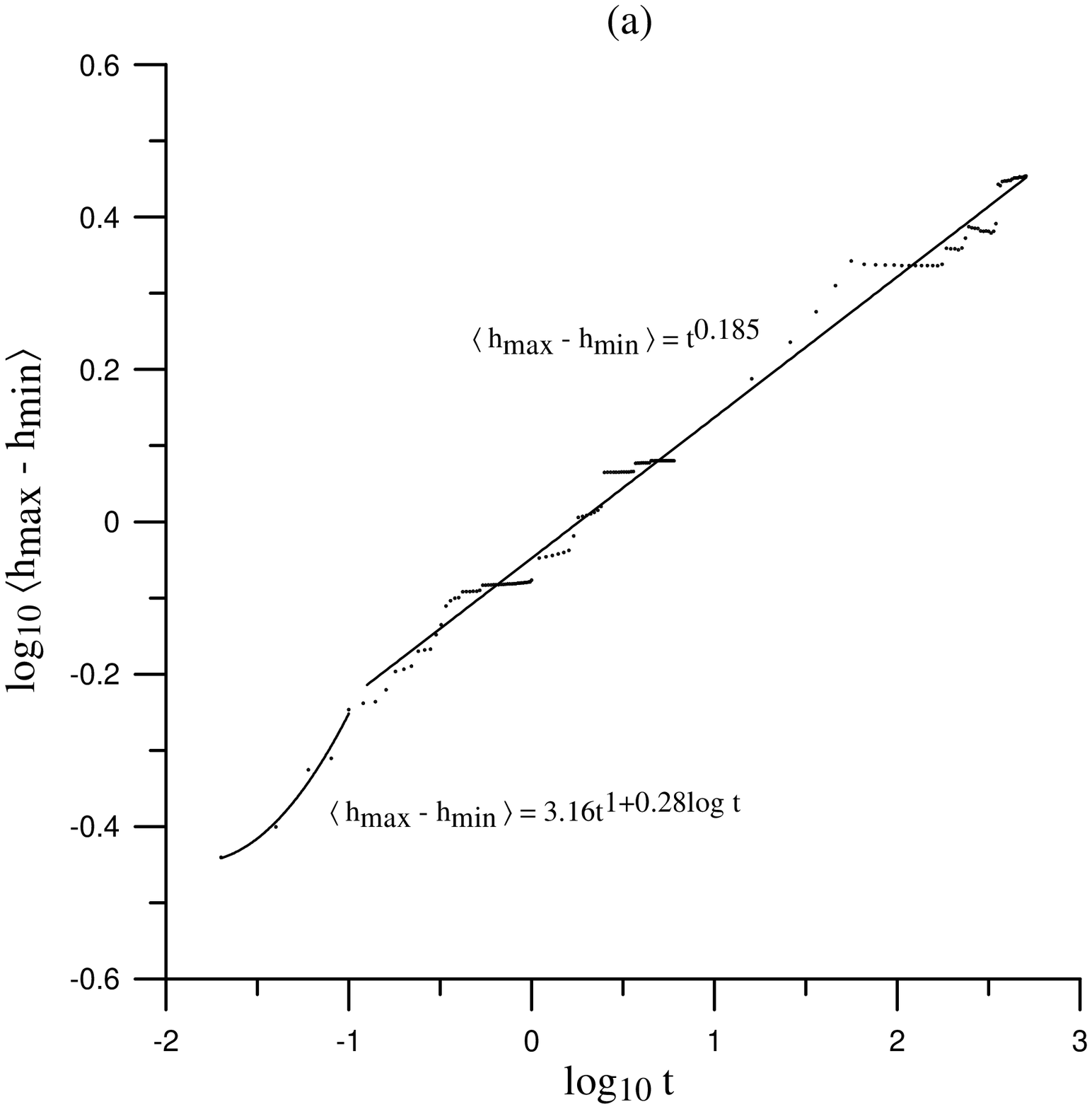}\includegraphics[width=3.0in]{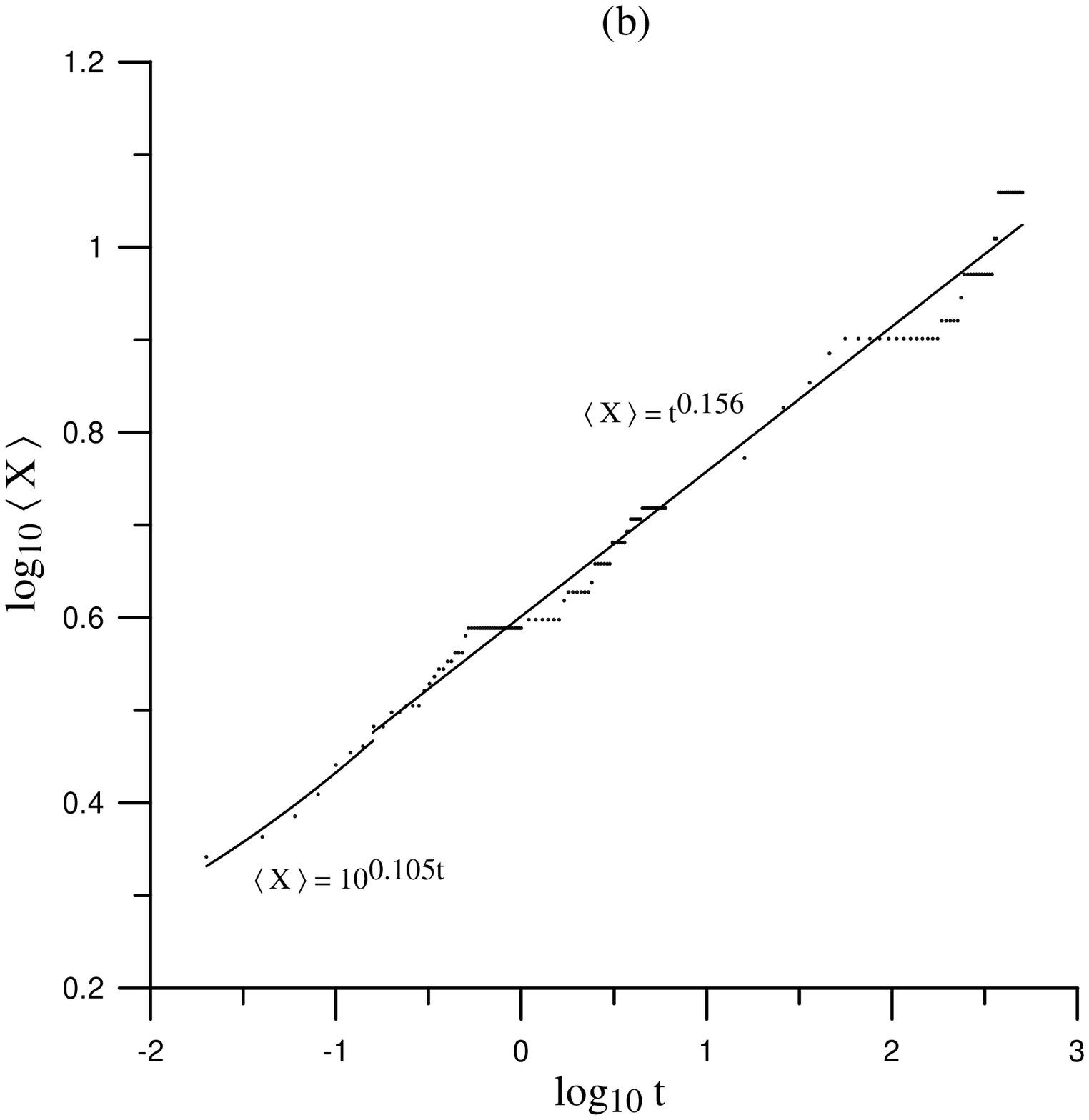}
\vspace{-1.0cm}
\caption{Wetting films, horizontal electric field, anisotropic mobility: (a) Log-log plot of the averaged (over ten realizations) maximum hill height; (b) Log-log plot of averaged 
kink-kink (valley-valley) distance. Dots: computed data, lines: fitting curves. $A=71,\ h_0=10,\ G=2$.}
\label{Fig_coarsening_A=71}
\end{figure}

\begin{figure}[h]
\vspace{-1.2cm}
\centering
\includegraphics[width=3.0in]{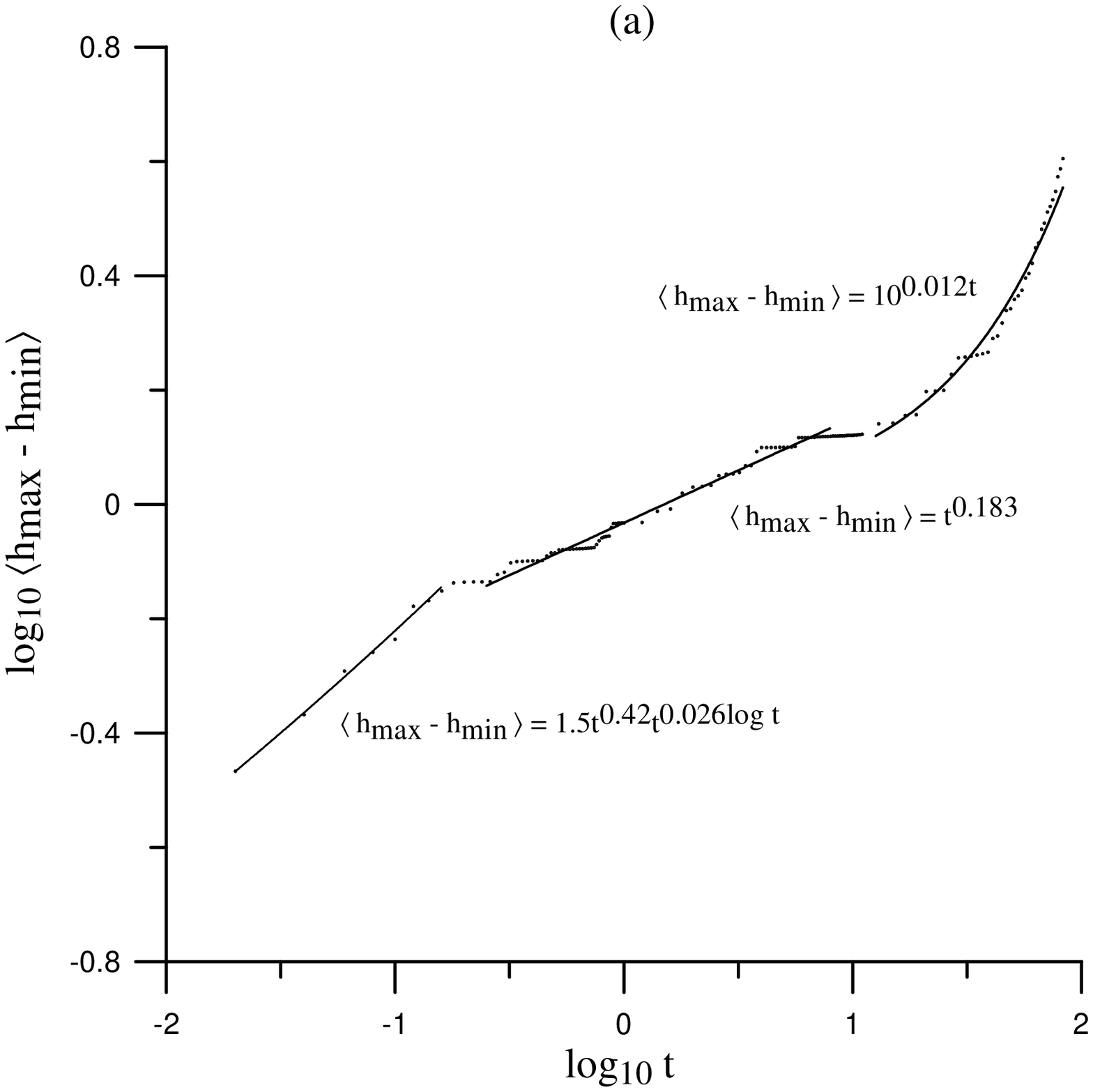}\includegraphics[width=3.0in]{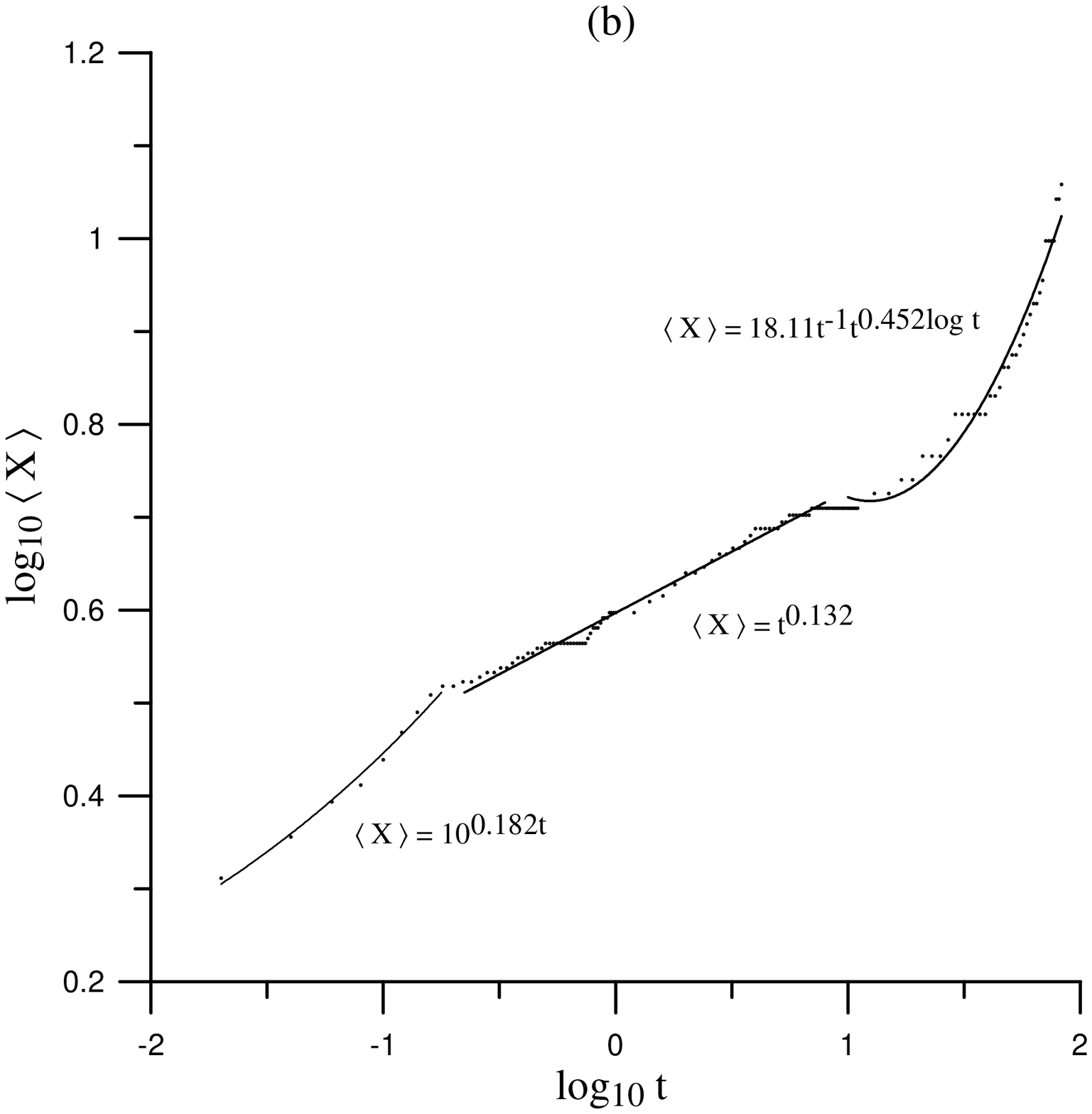}
\vspace{-1.0cm}
\caption{Same as Fig. \ref{Fig_coarsening_A=71}, for $A=71,\ h_0=6,\ G=2$.}
\label{Fig_coarsening_A71_G2_h0_6}
\end{figure}

\begin{figure}[h]
\vspace{-1.2cm}
\centering
\includegraphics[width=3.0in]{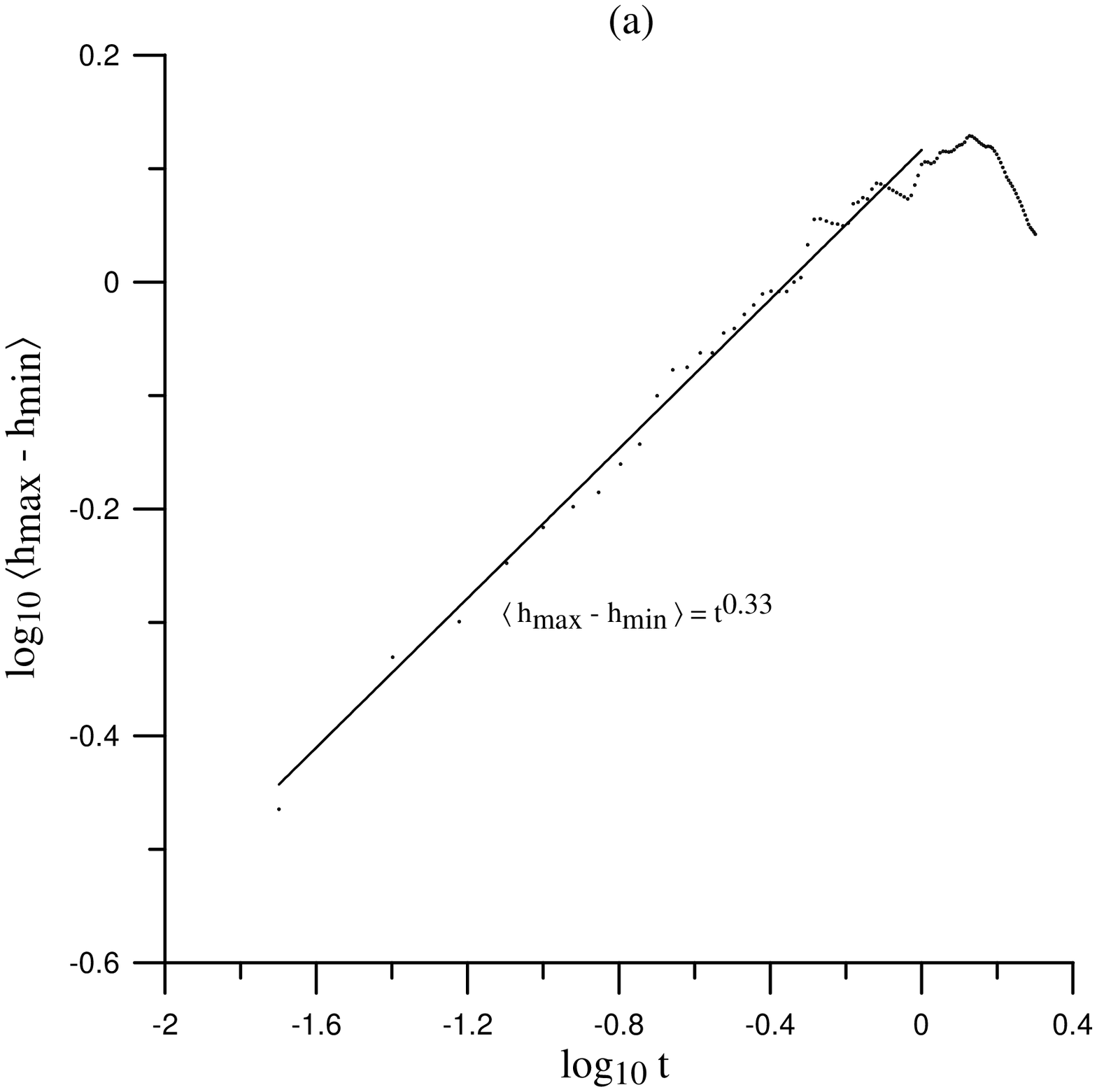}\includegraphics[width=3.0in]{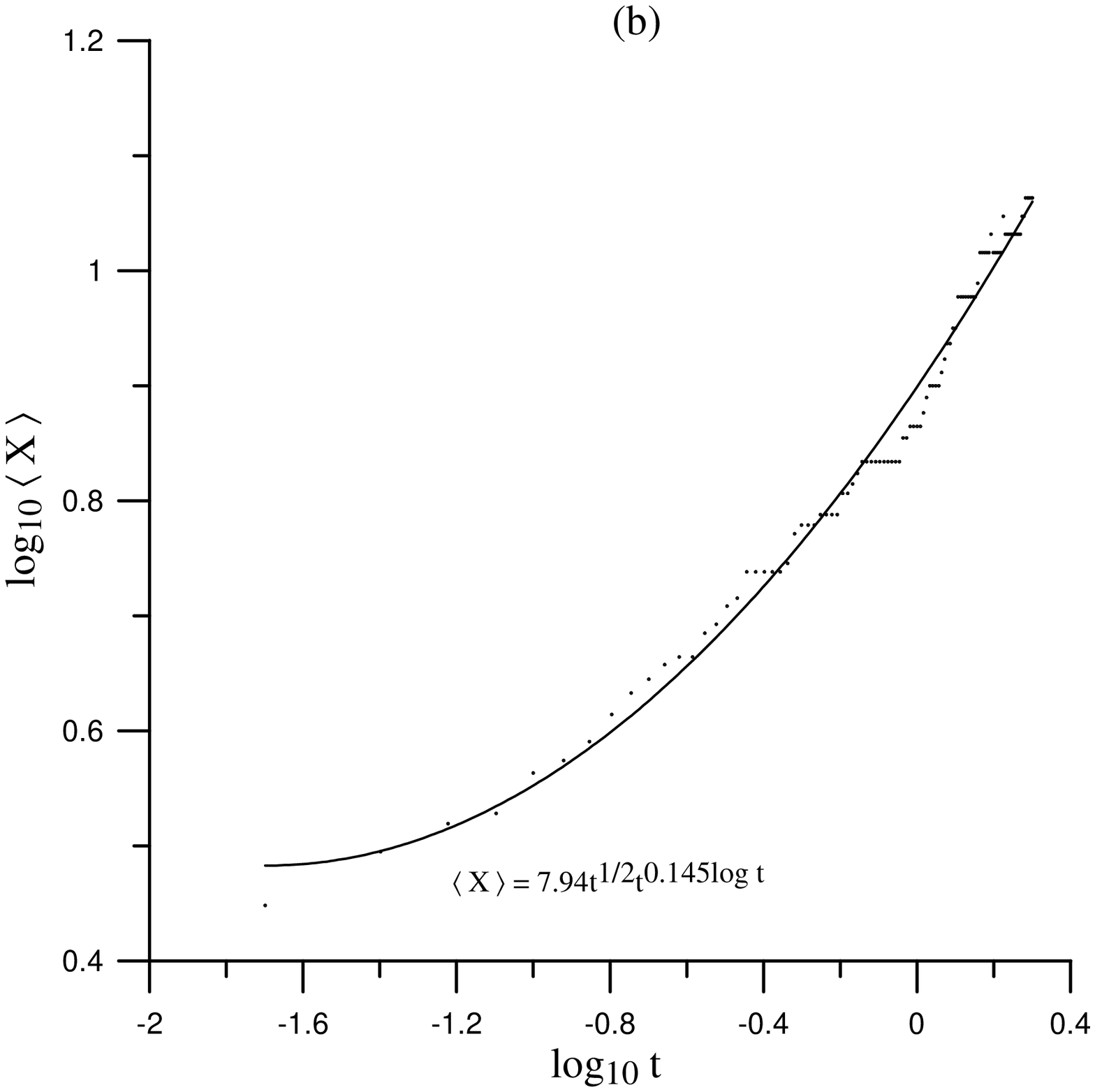}
\vspace{-1.0cm}
\caption{Same as Figs. \ref{Fig_coarsening_A=71} and \ref{Fig_coarsening_A71_G2_h0_6}, for $A=71,\ h_0=6,\ G=500$.}
\label{Fig_coarsening_A71_G500_h0_6}
\end{figure}

We employed fully nonlinear simulations based on parametric equations for determination of the coarsening laws at increasing electric field strength and at variable wetting strength (characterized by values of the parameters 
$h_0$ and $G$). Computations were performed on the domain $0\le x\le 20\lambda_{max}$ with periodic boundary conditions; all runs were terminated after the surface evolved into a large-amplitude 
hill-and-valley structure
with 3-5 hills. Unless wetting is strong, i.e. $h_0$ is small and $G$ is large - the case that is discussed in more detail below - the slopes of the hills are constant 24$^\circ$ during coarsening, except for the 
short initial period.  Figs. \ref{Fig_coarsening_A=71} - \ref{Fig_coarsening_A71_G500_h0_6} are the log-log plots of the 
averaged, over ten realizations, maximum surface amplitude $h_{max}-h_{min}$ and the averaged mean horizontal distance $X$ between valleys (kinks) vs. time. 

\textit{Remark 4.} We also attempted to compute coarsening dynamics resulting from the small-slope Eq. (\ref{h-eq-ssa-hor}). While the hill-and-valley structure does emerge and coarsens with time,
the characteristic constant hill slope is nearly 90$^\circ$. This suggests that additional nonlinear terms must be retained in Eq. (\ref{h-eq-ssa-hor}) for predictive computations and raises the
question what terms must be retained. We had not tried to obtain an adequate nonlinear small-slope model, leaving this agenda to future research.

Of course, increasing the field strength results in faster coarsening, as the times needed for coarsening into a ``final" structure
are $10^3$, $10$, and $1$ for field strengths $A=71,\ 710$ and $7100$, respectively. These values also 
point out the decrease of the rate of change of a coarsening rate with the increasing field strength. 
Other than that, the coarsening laws are similar for the three tested field strengths. 
Fits to the data in the case $A=71$ are shown in Figs. \ref{Fig_coarsening_A=71}(a,b). At small times coarsening is fast (exponential, see Fig. \ref{Fig_coarsening_A=71}(b)), then it changes to a slower power
law with the exponent in the range $\sim 0.1\div 0.2$. (Since the values of the amplitude's logarithm are negative, we were unable to fit the exponential law to the data in Fig. \ref{Fig_coarsening_A=71}(a), thus we fitted the quadratic, which results in the $t^{\epsilon_1+\epsilon_2\log_{10}t}$-type law.)

Remarkably, when $h_0$ is decreased from 10 to 6 a very different coarsening behaviors emerge, see Figs. \ref{Fig_coarsening_A71_G2_h0_6}(a,b) and
\ref{Fig_coarsening_A71_G500_h0_6}(a,b). At $G=2$, after the period of slow power-law coarsening at intermediate times, the coarsening accelerates to exponential coarsening or $t^{\epsilon_1+\epsilon_2\log_{10}t}$ type coarsening at late times (Figs. \ref{Fig_coarsening_A71_G2_h0_6}(a,b)). (For reference, the 
linear stability in the case $h_0=6$ is shown in Figs.  \ref{Fig_kc_vs_h0_A_G}(c) and \ref{Fig_omegamax_vs_h0_A_G}(c) by the dash-dotted line.) At $G=500$ we did not output a sufficient
number of data points in the beginning of the computation, but it is expected that initial coarsening is still faster than the power law. The amplitude starts to decrease towards the end of the simulation, and the kink-kink distance coarsens fast on the entire time interval.\footnote{We also mention that this accelerated coarsening dynamics is not reproduced by the small-slope Eq. \ref{h-eq-ssa-hor}, as can be expected from the \textit{Remark 4}. While there is some evidence of acceleration, this computation breaks down too fast.} 

\begin{figure}[h]
\vspace{-1.2cm}
\centering
\includegraphics[width=3.0in]{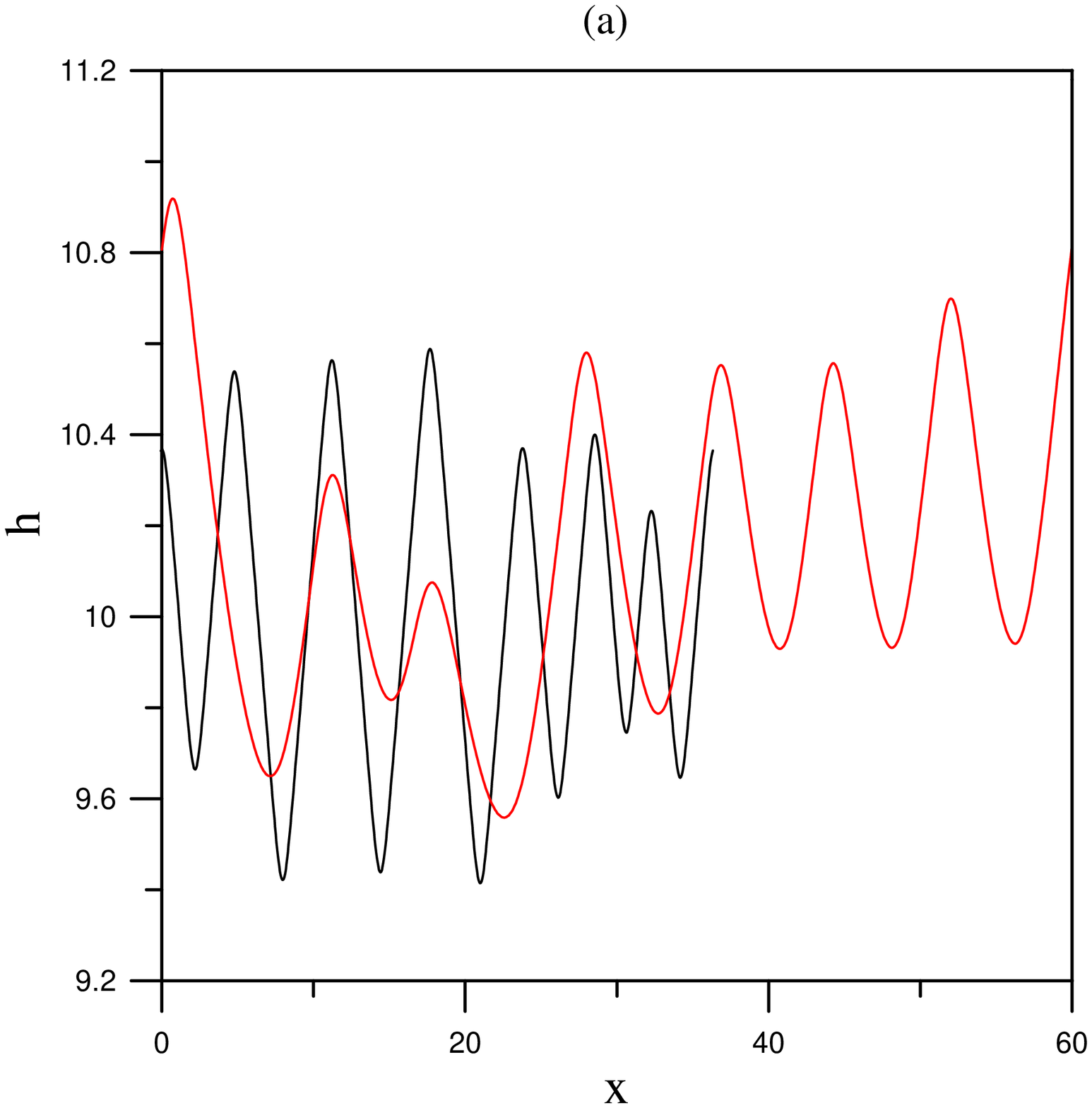}\includegraphics[width=3.0in]{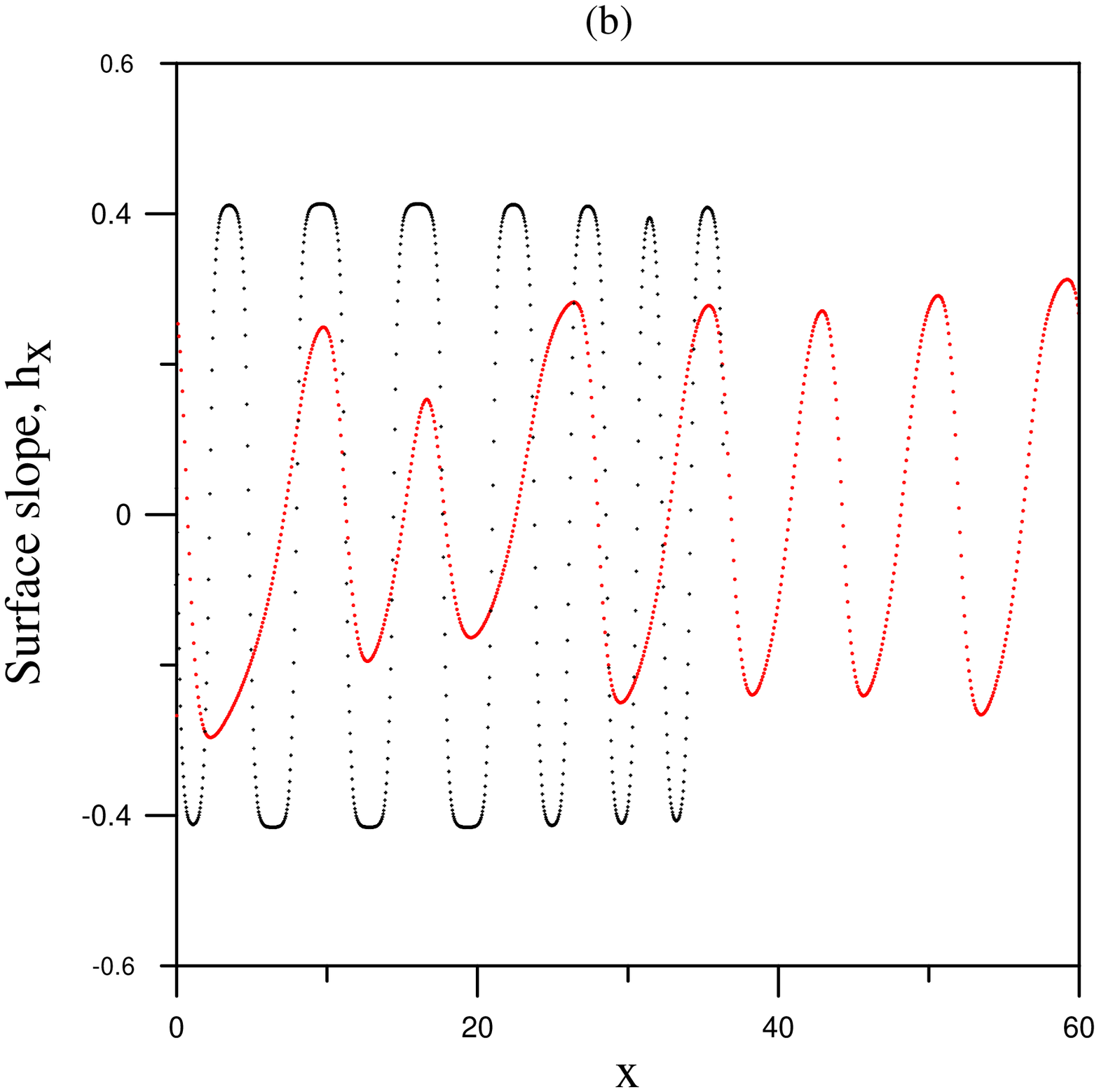}
\vspace{-1.0cm}
\caption{(Color online.) Wetting films, horizontal electric field, anisotropic mobility: (a) Surface height. Black line: $A=71,\ h_0=10,\ G=2$, red line: $A=71,\ h_0=6,\ G=500$; 
(b) Surface slope. Black crosses: $A=71,\ h_0=10,\ G=2$, red triangles: $A=71,\ h_0=6,\ G=500$. Note that the computational domains are of different length for the two presented cases, since
the $\lambda_{max}$ values differ.}
\label{Fig_comparison_A71_G500_h0_6_vs_A71_G2_h0_10}
\end{figure}

It seems certain that such unusual dynamics in the strongly wetting states emerges due to nonlinear "overdamping" of electromigration-induced faceting instability by the surface-substrate 
interaction force. Toward this end, in 
Figs. \ref{Fig_comparison_A71_G500_h0_6_vs_A71_G2_h0_10}(a,b) the surface shapes and surface slopes for $h_0=10,\ G=2$ and $h_0=6,\ G=500$ cases are compared at the time when
seven hills formed on the surface. In the former case the hills are steep,
have rather uniform height and their slopes are almost straight lines. In the latter case they are more irregular, "rounded", and the average height is smaller. 
It is reasonable to expect that in the opposite case of non-wetting films, the surface-substrate interaction will instead ``sharpen up" the hills, i.e. increase their slopes and make the surface structures
appear more spatially and temporally uniform.

Finally, we remark that the discussed coarsening laws for strong wetting cases also are qualitatively different from the laws governing coarsening in the absence of wetting, but in the presence of
deposition, attachment-detachment, strong surface energy anisotropy, and interface kinetics \cite{OL,SGDNV}. Indeed, only the coarsening exponents shown in Fig. \ref{Fig_coarsening_A=71} (week wetting)
are within the same range (0.1-0.5) for nearly the whole computational time interval, as are the exponents computed in the cited papers.


\section{Conclusions}
\label{Conclusions}

In this paper the effects of electromigration and wetting on thin film morphologies are discussed, based on the continuum model of film surface dynamics.
It has been shown that wetting effect modifies significantly the stability properties of the film and the coarsening of electromigration-induced surface roughness.
Also it has been shown that the small-slope evolution equations that were employed in many studies of electromigration effects on surfaces, are often inadequate
for the description of strongly-nonlinear phases of the dynamics.
It is expected that the account of the surface energy anisotropy and the electric field non-locality (through the solution of the moving boundary value problem for the electric 
potential) will lead to uncovering of even more complicated behaviours.



\section*{Acknowledgements}
This paper was inspired by the undergraduate research project  ``An analysis of the electromigration of atoms along a crystal surface using the Method of Lines Approach" by Kurt Woods (MATH 498 Senior Seminar, WKU Department of Mathematics and Computer Science, Fall 2011).

\end{document}